\documentclass[fleqn,usenatbib]{mnras}

\usepackage{newtxtext,newtxmath}

\usepackage[T1]{fontenc}

\DeclareRobustCommand{\VAN}[3]{#2}
\let\VANthebibliography\thebibliography
\def\thebibliography{\DeclareRobustCommand{\VAN}[3]{##3}\VANthebibliography}

\usepackage{graphicx}	
\usepackage{amsmath}	

\usepackage{amssymb}	

\title[The X-ray transient XTE J1859+226]{A refined dynamical mass for the black hole in the X-ray transient XTE J1859+226}

\author[I. V. Yanes-Rizo et al.]
{I. V. Yanes-Rizo$^{1,2}$\thanks{E-mail: idairayr@iac.es},
M. A. P. Torres$^{1,2}$,
J. Casares$^{1,2}$,
S. E. Motta$^{3}$,
T. Mu{\~n}oz-Darias$^{1,2}$,
\newauthor P. Rodr\'\i guez-Gil$^{1,2}$,
M. Armas Padilla$^{1,2}$,
F. Jim\'enez-Ibarra$^{4}$,
P. G. Jonker$^{5,6}$,
J. Corral-Santana$^{7}$ and
\newauthor R. Fender$^{8,9}$
\\
$^{1}$Instituto de Astrof\'\i sica de Canarias, E-38205 La Laguna, Tenerife, Spain\\
$^{2}$Departamento de Astrof\'\i sica, Universidad de La Laguna, E-38206 La Laguna, Tenerife, Spain\\
$^{3}$Istituto Nazionale di Astrofisica, Osservatorio Astronomico di Brera, via E. Bianchi 46, I-23807 Merate (LC), Italy\\
$^{4}$Australian Astronomical Optics, Macquarie University, 105 Delhi Rd, North Ryde, NSA 2113, Australia\\
$^{5}$Department of Astrophysics\,/\,IMAPP, Radboud University, Heyendaalseweg 135, NL-6525 AJ Nijmegen\\
$^{6}$SRON, Netherlands Institute for Space Research, Niels Bohrweg 4, 2333 CA, Leiden, The Netherlands\\
$^{7}$European Southern Observatory, Alonso de Córdova 2107, Vitacura, Casilla 19001, Santiago de Chile, Chile\\
$^{8}$Department of Physics, University of Oxford, Denys Wilkinson Building, Keble Road, Oxford OX1 3RH, UK\\
$^{9}$Department of Astronomy, University of Cape Town, Private Bag X3, Rondebosch 7701, South Africa\\
}

\date{Accepted XXX. Received YYY; in original form ZZZ}

\pubyear{2022}

\begin{document}
\label{firstpage}
\pagerange{\pageref{firstpage}--\pageref{lastpage}}
\maketitle

\begin{abstract}
We present two contiguous nights of simultaneous time-resolved GTC spectroscopy and WHT photometry of the black hole X-ray transient XTE J1859+226, obtained in 2017 July during quiescence. Cross-correlation of the individual spectra against a late K-type spectral template enabled us to constrain the orbital period to $0.276 \pm 0.003$\,d and the radial velocity semi-amplitude of the donor star to $K_2 = 550 \pm 59$\,km s$^{-1}$. An ellipsoidal modulation is detected in the photometric $r$- and $i$-band light curves, although it is strongly contaminated by flickering activity. By exploiting correlations between the properties of the double-peaked H$\alpha$ emission-line profile and the binary parameters, we derived an orbital inclination of $66.6 \pm 4.3$\,deg, a refined $K_2 = 562 \pm 40$\,km s$^{-1}$ and mass ratio $q = M_2/M_1 = 0.07 \pm 0.01$. From these values we obtained an updated black hole mass of $M_1 = 7.8 \pm 1.9$\,M$_\odot$. An independent mass estimate based on X-ray timing agrees well with our value, which gives further support for the outburst QPO triplet being explained by the relativistic precession model. We also obtained a companion star mass $M_2 = 0.55 \pm 0.16$\,M$_\odot$, which is consistent with its K5-K7\,V spectral type.
\end{abstract}


\section{Introduction}

X-ray transients contain a neutron star or a black hole that accretes matter via an accretion disc from a Roche lobe overflow low-mass companion. The accretion discs undergo occasional outbursts typically lasting from a few months to about one year caused by mass transfer instabilities  \citep[e.g.][]{mcclintock2006}. These X-ray transients spend most of the time in a low luminosity quiescent state, when their optical emission frequently includes a significant contribution from the companion star. This offers the best opportunity to analyse the orbital motion of the companions and obtain dynamical information that eventually allows  the masses of the accreting compact objects to be constrained \citep[see][for a review]{casares2014}. Further, X-ray transients may provide important insights on the nature and formation of compact objects. For example, dynamical studies of these systems are being used over the past decades to build the black hole mass and space distributions \citep[e.g.][and references therein]{ozel2010, kreidberg2012, jonker2021}, that can be compared with predictions from supernova models and observations of gravitational wave sources \citep[e.g.][]{fryer2001, belczynski2021}. 

Reliable black hole masses and knowledge of the binary geometry are necessary for testing models of the physical processes responsible for the spectral and temporal properties displayed by these systems during the outburst and quiescent states. A relevant example is the detection of X-ray quasi-periodic oscillations (QPOs) during outburst, that can help us to probe the strong gravitational field near the black hole \citep{motta2014qpo, ingram2019}.

XTE\,J1859+226 (hereafter J1859) is a black hole X-ray transient discovered by the \emph{Rossi X-Ray Timing Explorer} (\emph{RXTE}) on 1999 October 9 \citep{wood1999}. The X-ray counterpart peaked at $\simeq 1.5$\,Crab eight days after its discovery \citep{focke2000}. Its X-ray spectral and temporal behaviour resembled other black hole X-ray transients \citep{dalfiume1999, markwardt1999}. The source was subsequently detected across the spectrum, from radio to gamma rays \citep{mccollough1999, pooley1999, brocksopp2002}. The optical counterpart of J1859 was an $R = 15.1$\,mag object that showed emission lines characteristic of X-ray transients during outburst \citep{garnavich1999, wagner1999}. The ultraviolet counterpart was used to constrain the reddening towards the source and model the evolution of the accretion disc spectrum along the outburst decay, including viscous heating and X-ray irradiation \citep{hynes2002}. J1859 showed three re-brightening during the decline to quiescence and remained active until the end of 2000 August, when its optical counterpart reached a quiescent magnitude of $R = 22.48 \pm 0.07$ and $V = 23.29 \pm 0.09$\,mag \citep{zurita2002}.

Photometric monitoring during outburst \citep{uemura1999, mcclintock2000} and quiescence \citep{zurita2002} revealed tentative orbital periods ranging from 6 to 19\,h. A combined analysis of optical spectra and photometry taken at different epochs during quiescence delivered an orbital period of $6.58 \pm 0.05$\,h, a radial velocity semi-amplitude of the donor star of $K_2 = 541 \pm 70$\,km s$^{-1}$ and a mass function of $4.5 \pm 0.6$\,M$_\odot$, confirming the black hole nature of J1859 \citep{Corral-Santana2011}. 

In this work, we present simultaneous optical spectroscopy and photometry of J1859. The main aim of this project was to provide the first determination of the binary inclination by modelling the ellipsoidal light curve after correcting for the accretion disc contamination using the spectroscopy. Unfortunately, this estimation was hampered by the presence of very strong photometric variability typical of the active quiescence state \citep[see][]{cantrell2008}. Nevertheless, we provide refined system parameters based on an alternative approach. 

The paper is organized as follows: the observations and their reduction are outlined in Section~\ref{sec:2}. The results are presented in Section~\ref{sec:3}, where we derive the orbital period and the radial velocity semi-amplitude of the donor star. In addition, we provide constraints on the binary mass ratio and the inclination from the analysis of the H$\alpha$ emission line properties. Finally, in Section~\ref{sec:4} we give refined masses for the black hole and the donor star and compare the black hole mass with a recent estimation obtained from X-ray timing analysis.

\section{Observations and data reduction}
\label{sec:2}
We obtained simultaneous time-resolved spectroscopy and photometry of J1859 on 2017 July 22 and 23 with the 10.4-m Gran Telescopio Canarias (GTC) and the 4.2-m William Herschel Telescope (WHT), respectively, at the Observatorio del Roque de los Muchachos on the island of La Palma. During the observations the airmass ranged between 1 and 2.8 under dark sky conditions. We measured the seeing from the WHT images and the spatial profiles of the GTC spectra. On the first night, the seeing was variable between $0.8-1.8$\,arcsec and deteriorated gradually on the second night from 1.1 at the beginning to 2.8\,arcsec towards the end of the night.

\subsection{Spectroscopy}
\label{sec:21} 
Intermediate resolution spectroscopy of J1859 was obtained with the OSIRIS spectrograph \citep{cepa2000} on the GTC. We used the R1000R grism and the Marconi CCD44--82 set to 2x2 binning mode to cover the spectral range $5100-10000$\,\AA\ with a 2.6 {\AA} pix$^{-1}$ dispersion. The slit width was selected depending on seeing conditions. In this way, ten, four and twelve 1895-s exposures were taken with slit widths of 0.8, 1.0 and 1.23\,arcsec, respectively, that delivered spectra with 6.5, 7.8 and 9.9\,\AA\ full-width at half-maximum (FWHM) resolution. The quoted FWHMs were measured from Gaussian fits to the [O\,{\sc i}] 5577.334\,\AA\ night-sky emission line. {\sc pyraf}\footnote{\url{https://github.com/iraf-community/pyraf}} was used for reduction, extraction and wavelength calibration of the data. This last task was performed with a two-piece cubic spline fit to 15-20 lines of combined Xe, Ne and HgAr comparison arc lamp exposures obtained at the beginning of each night. The residual root mean square (rms) scatter of the fit was always $<0.1$\,\AA, less than four per cent of the wavelength dispersion. We used the [O\,{\sc i}] 5577.334\,\AA\ sky line positions measured from Gaussian fits to correct for wavelength zero point shifts, that ranged between 3 and 100\,km s$^{-1}$. The extracted spectra have a mean continuum signal-to-noise ratio (SNR) of $\simeq 6$ in the $6000-6300$\,\AA\ wavelength region. We imported them to the {\sc molly}\footnote{\textsc{molly} was written by T.~R. Marsh and is available from \url{https://cygnus.astro.warwick.ac.uk/phsaap/software/molly/html/INDEX.html}.} package, where we applied the heliocentric correction and performed all subsequent analysis. The J1859 spectra were continuum normalised with three-knot spline fits after masking out the telluric bands and emission lines. 
For the analysis we used the 61 Cyg B (K7 V) and 61 Cyg A (K5 V) spectral type templates with a resolution of 5\,\AA, presented in \citet{Casares1995} and \citet{Corral-Santana2011}, respectively. In addition, we used seven late K-V and two M0-V 2.5\,\AA\ resolution spectra from the MILES library \citep{falcon2011}. These were selected on the basis of a [Fe/H] metallicity ranging between 0 and 0.3\,dex. The template spectra were continuum normalized and corrected for radial velocity and wavelength zero-point offsets by removing the velocity measured from Gaussian fits to the core of the H$\alpha$ absorption line. When necessary, we convolved the template spectra with a Gaussian function to match the average instrumental resolution of the J1859 OSIRIS data.

\subsection{Photometry}
We performed time-resolved photometry simultaneous to the GTC spectroscopy with the Auxiliary-port CAMera \citep[ACAM;][]{Benn2008} on the WHT. We took 78 pairs of 300-s SDSS $r$ and $i$ images that were bias subtracted and flat field corrected using {\sc pyraf}. We used the HiPERCAM\footnote{\url{https://github.com/HiPERCAM/hipercam}} data reduction pipeline to perform optimal aperture photometry \citep{Naylor1998} on J1859, five comparison stars reported in \citet{zurita2002} and two additional field stars with similar brightness to our target. The field was calibrated using the Pan-STARSS Data Release 2 catalogue \citep{chambers2016}. To allow comparison of our J1859 photometry with previous works, we calculated the $R$-band magnitudes using the transformation by Lupton (2005)\footnote{\url{http://classic.sdss.org/dr4/algorithms/sdssUBVRITransform.html}}: $R = r - 0.2936\,(r - i) - 0.1439$. This reproduced the $R$-band magnitudes of the field stars in \citet{zurita2002} within the uncertainties.

\section{Analysis and results}
\label{sec:3}

\subsection{Spectroscopy}
\subsubsection{Radial velocities and orbital period}\label{sec:rvel_porb}
Radial velocities were measured via cross-correlation of the twenty-six spectra against the templates described in Section 2.1. After several tests the cross-correlations were performed in the wavelength regions $5930-6280$, $6310-6510$ and $6620-6800$\,\AA\ using a maximum shift window of $\pm 1000$\,km s$^{-1}$. The cross-correlation functions (CCFs) derived with the 61 Cyg B (K7 V) template star were less noisy than those obtained with other templates. The individual CCFs were inspected and eight were rejected because they did not show clear peaks or they hit the window edges. Thus, in total, we obtained eighteen useful radial velocities from the cross-correlation analysis. In addition, we fitted a symmetric two-Gaussian model to the double-peaked H$\alpha$ emission line in the individual twenty-six spectra to obtain the centroid velocity of the profile. The two Gaussians were forced to have identical FWHM and peak intensities. In this case, we excluded five data points with large errors caused by the poor SNR ($<3.7$) of the spectra.

\begin{table}
	\centering
	\caption{Best-fit radial velocity parameters. Uncertainties are at 1-$\sigma$ confidence level.}
	\label{tab:ephem}
	\begin{tabular}{ccccc} 
		\hline
		$K_2$ & $T_0$   & $P$  & $\gamma$  \\
		(km s$^{-1}$) & $(\mathrm{HJD}-2457000)$ & (d) & (km s$^{-1}$) \\
		\hline
		$550 \pm 59$ & $957.593 \pm 0.005$ & $0.276 \pm 0.003$ & $115 \pm 42$ \\
		\hline
	\end{tabular}
\end{table}

\begin{figure}
	\includegraphics[width=\columnwidth]{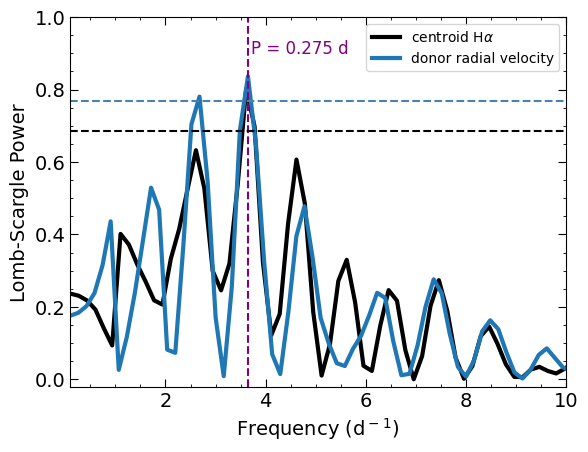}
    \caption{Lomb-Scargle periodograms from the companion star radial velocities (blue) and the H$\alpha$ emission-line centroid velocities (black). The horizontal blue and black dashed lines show the 99 per cent white noise significance level on the radial and centroid periodogram, respectively. The vertical dashed line marks the frequency at 3.64 cycle d$^{-1}$ (0.275-d period).}
    \label{fig:panel2_figure}
\end{figure}

\begin{figure}
	\includegraphics[width=\columnwidth]{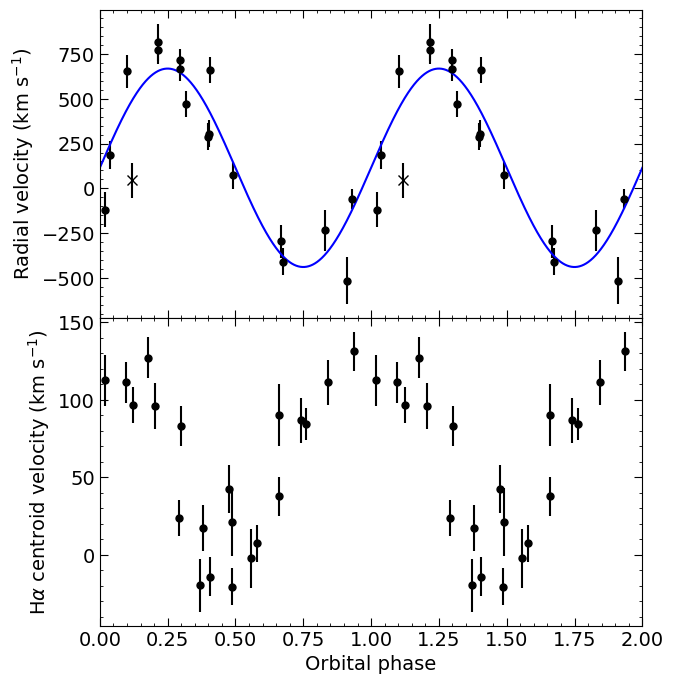}
    \caption{Radial velocity curve of the companion star absorption features (top) and H$\alpha$ emission-line centroid velocities (bottom) phase folded on the ephemeris provided in Table \ref{tab:ephem}. The blue solid line represents the best sine fit to the companion velocities. The data point marked with a cross is 4.5-$\sigma$ off from the best-fit model. The orbital cycle has been plotted twice for clarity.}
    \label{fig:panel3_figure}
\end{figure}

We computed Lomb-Scargle periodograms \citep{lomb76-1, scargle82-1} from the CCF radial velocities and the H$\alpha$ emission-line centroids (see Fig.~\ref{fig:panel2_figure}). In both cases, the frequency of the highest peak is found at 3.64\,cycle d$^{-1}$ (0.275-d period) and the second highest (which is a 1-d alias of the former) at 2.67\,cycle d$^{-1}$ (0.375-d period). The first peak is very likely the orbital period, as it is consistent with the period reported by \citet{Corral-Santana2011}. Furthermore, as shown in Fig.~\ref{fig:panel2_figure}, the 0.375\,d peak corresponding to the H$\alpha$ centroid velocity periodogram is below the 99 per cent white noise significance level. A bootstrap test has also been computed with 100 periodograms of the radial velocities and the H$\alpha$ centroids after randomly dropping three data points each time. We, respectively, find that in 80 and 90 per cent of the cases the highest peak is consistent with the 0.275 d periodicity. We subsequently performed sine fits of the form \begin{equation}
    v (t) = \gamma + K_2 \sin{\frac{2\pi(t-T_0)}{P}} 
    \label{eq:radialvel}    
\end{equation}
to the companion star radial velocities, where $v(t)$ is the measured radial velocity, $\gamma$ the systemic velocity, $t$ the observation Heliocentric Julian Date (HJD), $T_0$ the time of phase zero (corresponding to the inferior conjunction of the companion star) and $P$ the orbital period.

The sine wave fit, using $P=0.275$\,d as input for the period, gives a reduced $\chi^2 = 4.0$, with 13 degrees of freedom (dof). For comparison, the same fit for $P = 0.375$\,d resulted in a reduced $\chi^2 = 6.5$. We also fitted sine functions to the radial velocities derived from the H$\alpha$ emission-line centroids using either period. The curve folded on the 0.275-d period shows the least scatter of the two, with maximum and minimum velocity near orbital phases 0 and 0.5, respectively. The best-fit parameters for the CCF radial velocity curve of the companion star are shown in Table \ref{tab:ephem}, where the 1-$\sigma$ uncertainties were calculated after scaling the data uncertainties to obtain $\chi^2/\mathrm{dof} = 1$.

Fig.~\ref{fig:panel3_figure} shows the radial velocity curves of the companion star and the H$\alpha$ emission-line centroids folded on the ephemeris provided in Table~\ref{tab:ephem}.  Note that one of the data points (marked with a cross in the figure) was excluded from the fit because it is off by 4.5-$\sigma$ from the best-fit sine model. In any case, including this point does not have a significant effect in the fit. 

The radial velocity semi-amplitude of the companion star and the orbital period agree within 1-$\sigma$ with the values obtained by \citet{Corral-Santana2011}. However, note that our gamma velocity is significantly larger. We have checked the radial velocity of the K-type template used in \citet{Corral-Santana2011} and found a mistake in the flexure correction that produced an artificial shift in the systemic velocity reported in that work and that we corrected it.

\begin{figure*}
	\centering \includegraphics[height=8cm]{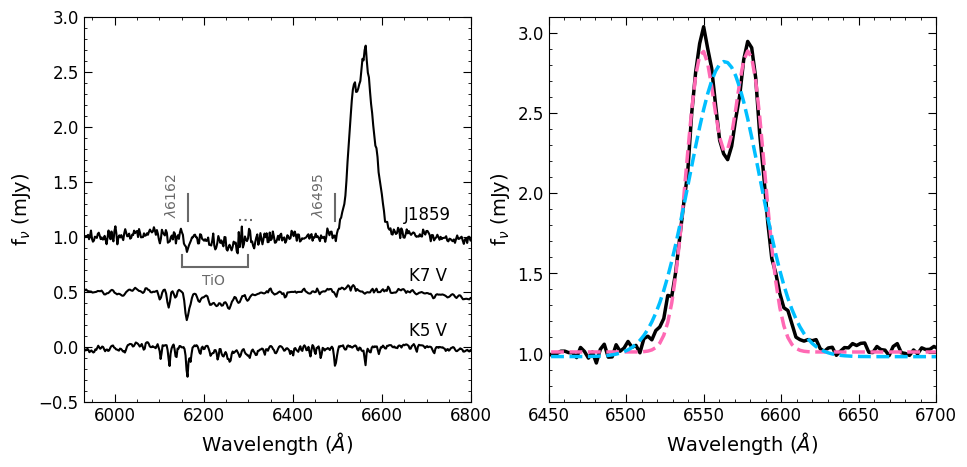}
    \caption{Left panel, from bottom to top: spectra of the K5\,V and K7\,V templates and the average spectrum of J1859 in the rest frame of the companion star. The location of expected strong photospheric absorptions for late K-type dwarfs such as Ca\,{\sc i} 6162\,\AA, the Fe\,{\sc i}+Ca\,{\sc i} blend at 6495\,\AA\ and the TiO band between these two features are marked. The wavelength interval masked out in the cross correlation analysis that is affected by strong telluric features and the residual of the [O\,{\sc i}] 6300\,\AA\ sky emission line is also traced as a horizontal dotted line. The spectra have been shifted vertically for display purposes. Right panel: the average H$\alpha$ emission line profile at the heliocentric rest frame is shown in black with blue and pink dashed lines representing the best-fit single- and double-Gaussian models, respectively.}
    \label{fig:panel6_figure}
\end{figure*}

As an independent test of the companion's radial velocity semi-amplitude, we used the empirical $K_2$--FWHM relationship for the H$\alpha$ emission line in quiescent X-ray transients \citep{casares2015}:
\begin{equation}
    K_2 = 0.233(13) \times \rm{FWHM(H}\alpha)~.
	\label{eq:K2}
\end{equation}
We measured the FWHM of the H$\alpha$ emission line in J1859 and its uncertainty from the mean and standard deviation of the distribution of values provided by single Gaussian fits to the individual emission profiles, and obtain ${\rm FWHM} = 2462 \pm 118$\,km s$^{-1}$. This yields $K_2 = 574 \pm 60$\,km s$^{-1}$, which is fully consistent with the value obtained from the radial velocity curve.

\subsubsection{Spectral type of the companion star}
\label{sec:sptype}
To constrain the spectral type of the companion star in J1859 we first shifted the individual spectra of J1859 to the rest frame of the companion star by removing the CCF radial velocity curve. Next, we averaged the Doppler-shifted spectra using individual weights proportional to the inverse of the variance in order to optimise the SNR of the final spectrum. The left panel of Fig.~\ref{fig:panel6_figure} shows the resulting Doppler-corrected, average spectrum of J1859 and the spectra of the K5 and K7\,V templates that were used for the cross correlation. Due to the large contribution of the accretion disc to the total flux and the limited SNR of the data, the stellar absorption lines are barely detectable except for the Ca\,{\sc i} 6162\,\AA\ line \citep{allen1995} and the TiO band at $6150-6300$\,\AA. The detection of this molecular band is consistent with a spectral type later than K5 \citep[e.g.][]{jacoby1984}, while its depth (lower than that of Ca\,{\sc i} 6162\,\AA) is indicative of a spectral type earlier than M1 \citep[e.g.][]{zhong2015}. Note that the expected Fe\,{\sc i}+Ca\,{\sc i} blend at 6495\,\AA\ characteristic of K-type stars \citep{horne1986} is poorly detected as it is contaminated by the H$\alpha$ emission line. Given that the M-type templates that we used gave poor cross correlation results, we favour a K5-7\,V spectral type for the companion star in line with earlier classification attempts \citep{Corral-Santana2011}. We also applied the optimal subtraction technique described in \citet{marsh1994}. It searches for the lowest residual spectrum after subtracting a set of broadened and scaled spectral templates to the average Doppler-corrected spectrum of the target. However, the optimal subtraction of the K5 V, K7 V and MILES templates from the Doppler corrected average spectrum did not result in a clear spectral classification. While the above technique also serves to establish the projected rotational broadening of the absorption lines, the spectral resolution of our data was insufficient for this purpose. From the values of $K_2$ and $q$ obtained in this work, we predict a broadening of $110 \pm 10$\,km s$^{-1}$.

\subsection{Photometry}
\begin{figure}
	\includegraphics[width=\columnwidth]{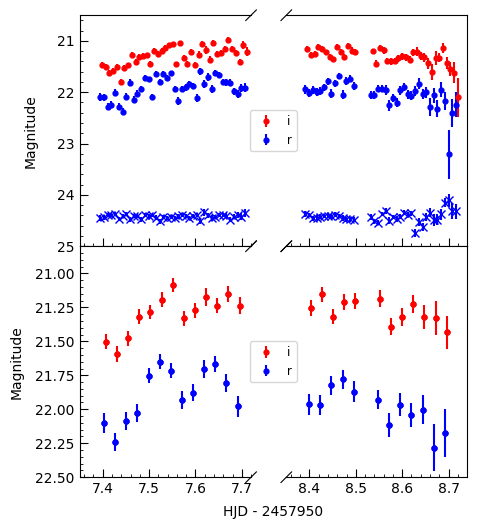}
    \caption{Top: WHT/ACAM $r$ (blue points) and $i$ (red points) light curves from data obtained on 2017 July 22 and 23. The blue crosses represent the $r$-band magnitudes of a field star with similar brightness to J1859. It has been shifted vertically by 2.5\,mag for display purposes. Bottom: smoothed light curves of J1859 obtained by averaging the data points into 30-min time bins. The error represents the rms of the magnitudes.}
    \label{fig:panel4_figure}
\end{figure}

The $r$- and $i$-band light curves of J1859 are shown in Fig.~\ref{fig:panel4_figure}. For comparison purposes, we also plot the light curve of a field star with similar magnitude ($r = 21.92 \pm 0.04$, $i = 20.87 \pm 0.03$) to J1859.
The light curves show strong aperiodic short-term variability (flickering), that was also detected in the 2010 photometry reported in \citep{Corral-Santana2011}. This flickering has typical amplitudes in the range $0.06-0.6$\,mag in X-ray transients, and can sometimes dominate over the ellipsoidal modulation caused by the tidally-distorted companion star \citep[see e.g.][]{zurita2003}. In an attempt to unveil the underlying ellipsoidal variability, we have constructed three-point ($\simeq 30$-min time resolution) binned light curves (bottom panel in Fig.~\ref{fig:panel4_figure}). The resulting binned light curves for the night of 2017 July 22 show two maxima and two minima, consistent with what is expected from an ellipsoidal modulation. This is less clear on the second night, most likely due to the very poor seeing conditions that affected its second part, where the rms variability of the faint comparison field star was up to 0.1\,mag. Therefore, we restrict our analysis to the data obtained on 2017 July 22 and the first part of the night of July 23 ($\mathrm{HJD}$ from  2457957.39 to 2457958.53). Clear variability with an rms of 0.2\,mag is identified in J1859, which is a factor of 5 larger than that for comparison field stars of similar brightness. We establish mean magnitudes for J1859 of $r= 21.9 \pm 0.2$ and $i = 21.3 \pm 0.2$\,mag. Following the transformations of Lupton (2005), this corresponds to a mean $R$-band magnitude of $21.6 \pm 0.3$, which compares well with the magnitude reported in \citet{Corral-Santana2011}. Therefore, we conclude that the target was brighter during the course of our observations than in the $R = 22.48 \pm 0.07$ passive (lowest level flickering) quiescent state observed in 2000 August \citep{zurita2002}.

\begin{figure}
	\includegraphics[width=\columnwidth]{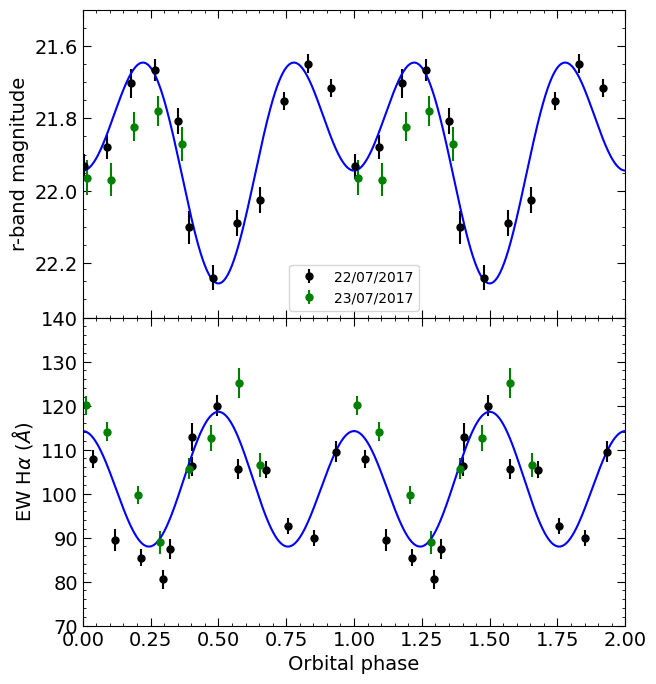}
    \caption{Top: phase folded $r$-band light curve of J1859 using the ephemeris provided in Table \ref{tab:ephem}. The light curve shows a double-humped morphology with the deeper minimum at orbital phase 0.5, characteristic of an ellipsoidal modulation. Bottom: EWs of the H$\alpha$ emission line versus orbital phase. Their orbital evolution is in anti-phase with the above continuum variations. Double sine-wave fits (solid blue lines) are included in both plots to guide the reader's eye for the global trend. One orbital cycle is repeated for clarity.}
    \label{fig:panel5_figure}
\end{figure}

Fig.~\ref{fig:panel5_figure} presents the binned $r$-band photometric light curve folded on the spectroscopic ephemeris (Table~\ref{tab:ephem}). In this photometric band the ellipsoidal variability is more clear than in the $i$-band. As can be seen, the photometric minima occur at phases 0.0 and 0.5, with the deeper minimum at phase 0.5, as expected for an ellipsoidal modulation with gravity darkening \citep{greene2001, shahbaz2003}. As further evidence for the presence of an ellipsoidal modulation in our data, we also display in Fig.~\ref{fig:panel5_figure} (bottom panel) the orbital evolution of the H$\alpha$ emission equivalent width (EW), folded on the spectroscopic ephemeris. We see a ‘reflection' of the double-humped $r$-band modulation, with maxima at phases 0 and 0.5, as predicted if this were dominated by continuum variations caused by the tidal distortion of the companion star. Similar signs of EW ellipsoidal variations have been reported for A0620--00 \citep[e.g.][]{marsh1994, neilsen2008}.

\subsection{The binary mass ratio and inclination}
A companion star to black hole mass ratio $q = M_2/M_1 = 0.049_{-0.012}^{+0.023}$ for J1859 was previously calculated by \citet{Casares2016} from the FWHM of the average H$\alpha$ emission profile presented in \citet{Corral-Santana2011}. Thanks to our more extended data set, we are able to revisit this parameter by fitting our average H$\alpha$ profile with Gaussian templates and using the correlation:
\begin{equation}
    \log q = -6.88(0.52) - 23.2(2.0) \log\left(\frac{DP}{FWHM}\right)~,
	\label{eq:massratio}
\end{equation}
where $DP$ and $FWHM$ are the double peak separation and the FWHM of the profile, respectively. Following \citet{Casares2016}, we performed fits to the average H$\alpha$ profile using both a single and a symmetric two-Gaussian model (see right panel of Fig.~\ref{fig:panel6_figure}), with the components of the latter having equal height and FWHM. Before conducting the fits, these models were degraded to the 8.0 {\AA} resolution of our average spectrum. This was achieved by applying the same weights used for every spectrum to their specific instrumental resolution. In order to evaluate the 1-$\sigma$ uncertainty in $q$, we adopted a Monte Carlo approach. We drew 10\,000 random values from normal distributions centred on the mean value and with a FWHM equivalent to the 1-$\sigma$ uncertainties of the measurements in play. We obtained $DP = 1382 \pm 5$\,km s$^{-1}$ and $FWHM = 2442 \pm 12$\,km s$^{-1}$, which yield a mass ratio  $q = 0.07 \pm 0.01$. Our refined value is consistent to within 1-$\sigma$ of the estimate presented in \citet{Casares2016}.

The ellipsoidal modulation is clearly affected by flickering, so its amplitude cannot be used to estimate the orbital inclination, $i$, through light curve modelling. Instead, we employed the correlation between $i$ and the depth of the trough, $T$, in between the two peaks of the H$\alpha$ emission profile by \citet{casares2022}. The correlation, established from the average H$\alpha$ profiles of a sample of seven X-ray transients in quiescence with orbital inclinations ranging 35-75\,deg, is given by:
\begin{equation}
    i \,(\rm{deg}) = 93.5(6.5)\,T + 23.7(2.5)~, 
	\label{eq:inclination}
\end{equation}
with $T$ given by 
\begin{equation}
    T = 1 - 2 ^{1-\left(\frac{DP}{W}\right)^2}~,
	\label{eq:depth}
\end{equation}
where $W$ is the FWHM of the symmetric two-Gaussian model fit to the H$\alpha$ emission profile. We obtained $W = 1011 \pm 6$\,km s$^{-1}$ from the two-Gaussian fit reported at the start of this section. This results in $T = 0.47 \pm 0.01$. Following \citet{casares2022}, we have applied a small +0.01 systematic shift to the $T$ value in order to correct for instrument resolution degradation. This value was obtained from equation A1 in that paper, using an equivalent resolution $\Delta = 0.39$ \citep[see details in Appendix A of][]{casares2022}. In order to account for epoch-to-epoch changes in the T measurement, we repeated its calculation on the average H$\alpha$ emission profile obtained from the 2010 spectroscopy \citep{Corral-Santana2011}. We derive $W = 985 \pm 11$\,km s$^{-1}$ and $DP = 1312 \pm 8$\,km s$^{-1}$ that result in $\Delta = 0.33$ and a value of $T = 0.43 \pm 0.02$ after applying a shift of +0.01 calculated to compensate for the 5 {\AA} resolution of the data. Taking the weighted mean of the depth trough values, we obtain $T = 0.46 \pm 0.02$, yielding an orbital inclination for J1859 of:
$$ i = 66.6 \pm 4.3 \deg~,$$ 
where the uncertainty on $i$ has been calculated through a Monte Carlo simulation with 10 000 realizations. 

The binary inclination obtained from equation (\ref{eq:inclination}) lies within the $40-70$\,deg limits set by the unequal minima of the passive state ellipsoidal modulation and the lack of X-ray eclipses during outburst, and is also within 60 and 70\,deg found from modelling the ellipsoidal variability with null and 28 per cent disc contribution, respectively \citep{zurita2002, Corral-Santana2011}. Incidentally, our result agrees well with the $\simeq 60$\,deg inclination adopted in the study of the spectral properties of J1859 by \citet{hynes2002}, \citet{munoz2013}, \citet{motta2014qpo}, \citet{nandi2018} and \citet{kimura2019}, as well as the black hole mass distribution study by \citet{kreidberg2012}. 

\section{Discussion}
\label{sec:4}
Our phase-binned $r$-band light curve of J1859 shows a maximum peak-to-peak amplitude of $\simeq\,0.6$ mag, which is larger than expected for an ellipsoidal modulation even at maximum inclination $i = 90$\,deg. We used the \textsc{xrbinary} code\footnote{Software developed by E.~L. Robinson, see \url{http://www.as.utexas.edu/~elr/Robinson/XRbinary.pdf} for further details.} to predict an ellipsoidal modulation with a peak-to-peak amplitude of $<0.39$\,mag in the $R$-band adopting the orbital parameters obtained in this work, a Roche-lobe filling companion with $T_{\rm{eff}} = 4100$\,K and a null disc contribution to the optical light. In fact, the light curve observed during deep quiescence shows a potential peak-to-peak amplitude of $\simeq 0.4$\,mag \citep{zurita2002, Corral-Santana2011}. The larger amplitude observed in our phase-binned data can not be due to ellipsoidal variability and must be caused by flickering activity, which makes a derivation of a binary inclination from light curve modelling unreliable. However, we have been able to obtain a value for the inclination from the depth of the H$\alpha$ double peak trough, which has been calibrated against a sample of X-ray transients with robust inclination determinations \citep{casares2022}.

In this work, we have derived $K_2$ by two different methods: a direct fit of the radial velocity curve and, indirectly, from the $K_2-\mathrm{FWHM}$ correlation of \cite{casares2015}. We can combine the values provided by these two independent methods to obtain a more precise $K_2 = 562 \pm 40$\,km s$^{-1}$. This, together with the orbital period ($0.276 \pm 0.003$\,d), yields a mass function $f(M_1) = 5.2 \pm 1.1$\,M$_\odot$. The masses of the black hole and its companion star can be obtained from:
\begin{equation}
    M_1 = \frac{f(M_1)(1+q)^2}{\sin^3{i}}; \; \; \; \; \; M_2 = qM_1~.
    \label{eq:massfun}
\end{equation}
Using our refined values $i = 66.6 \pm 4.3$\,deg and $q = 0.07 \pm 0.01$ results in $M_1 = 7.8 \pm 1.9$ and $M_2 = 0.55 \pm 0.16$\,M$_\odot$\footnote{If we restrict ourselves to the $K_2$ value obtained from the radial velocity curve alone, the black hole and the companion star masses would be $7.4 \pm 2.5$ and $0.52 \pm 0.20$\,M$_\odot$, respectively.} at 1-$\sigma$ confidence level. The black hole mass falls near the peak of the Galactic black hole mass distribution \citep{ozel2010, farr2011}, while the companion star mass is consistent with that of a K5-7 dwarf \citep{pecaut2013}. Following \citet{casares2022}, the black hole mass can also be estimated with a 20 per cent uncertainty from the intrinsic width of the double peak profile ($W$) and the orbital period $P_{\rm{orb}}$ through the equation: 
\begin{equation}
    M_1^* = 3.45\times10^{-8} P_{\rm{orb}} \left(\frac{0.63W+145}{0.84}\right)^3   \rm{M_\odot}~,
	\label{eq:mass}
\end{equation}
with $P_{\rm{orb}}$ expressed in days. The resulting black hole mass is $M_1^* = 7.7 \pm 1.5$\,M$_\odot$, which agrees within the uncertainty with the value provided by equation~(\ref{eq:massfun}).  

\subsection{Comparing the black hole mass with an X-ray timing determination}
There are new exciting prospects for black hole mass and spin determination with X-ray timing analysis. Peaked features in the Fourier power spectra (QPOs) with centroid frequencies from mHz to a thousand Hz have been observed in many black hole and neutron star X-ray binaries \citep{vanderklis2006, motta2016}. One interpretation links the three frequencies observed to fundamental frequencies of the motion of matter in the strong field regime as predicted by general relativity, i.e. (from slower to faster) Lense-Thirring, periastron precession and Keplerian frequencies. Such frequencies are related to blobs of matter orbiting the compact object \citep{stella1999, motta2014qpo}. In the case of a black hole X-ray binary, the functional form of the relativistic precession model (RPM) frequencies around a spinning (Kerr) black hole depends solely on the mass and spin of the compact object and the radius at which the orbiting matter produces the QPOs. If the three QPOs are observed simultaneously, one can assume that all three frequencies correspond to the same distance (radius) from the black hole. Thus, the system of equations associated with the three QPO centroid frequencies can be solved exactly \citep[e.g.][]{ingram2014}. The application of this technique to the black hole X-ray transient GRO\,J1655--40 resulted in a black hole mass $M_1 (\mathrm{J1655}) = 5.31 \pm 0.07$\,M$_\odot$ \citep{motta2014J1655}, which is fully consistent with the dynamical mass determined through accurate optical spectrophotometric observations \citep{beer2002}. In an accompanying paper, Motta et al. (accepted) report the identification of a QPO triplet in J1859 from which the RPM yields $M_1 = 7.85 \pm 0.46$\,M$_\odot$ and an adimensional spin value of $0.149 \pm 0.005$. Our refined dynamical mass is in remarkable agreement with this QPO mass and, thereby, further supports the validity of the RPM model. 

\section{Conclusions}
In this work we present simultaneous time-resolved photometry and spectroscopy of the X-ray transient J1859 in quiescence. The radial velocities of both the companion star and the H$\alpha$ emission line are modulated with the orbital period. From the former we derive an orbital period  $0.276 \pm 0.003$\, d and a radial velocity semi-amplitude of the companion star $K_2 = 550 \pm 59$\,km s$^{-1}$.

Our photometric light curves are strongly contaminated by flickering and can not be used to infer the binary inclination through ellipsoidal fits. Instead, we exploit a scaling with the H$\alpha$ double peak trough and derive $i = 66.6 \pm 4.3$\,deg. We also present a refined $K_2 = 562 \pm 40$\,km s$^{-1}$ and a mass ratio $q = 0.07 \pm 0.01$. Our new parameters imply a black hole mass function of $f(M_1) = 5.2 \pm 1.1$\,M$_\odot$ and refined black hole and companion star masses of $M_1 = 7.8 \pm 1.9$ and $M_2 = 0.55 \pm 0.16$\,M$_\odot$ at 1-$\sigma$ confidence level, respectively. This dynamical black hole mass is in remarkable agreement with the mass derived from the QPO triplet observed in X-ray data during outburst. This provides extra support to the relativistic precession model.

More optical observations under better weather conditions are required to further refine the fundamental parameters of J1859. A reduction in the $K_2$ uncertainty to $\sim 10$\,km s$^{-1}$ would decrease the uncertainties in the masses by a factor of two.

\section*{Acknowledgements}
We thank Ayoze \'Alvarez Hern\'andez for practical help during the data analysis. The {\sc molly} software by Tom Marsh is gratefully acknowledged. This article is based on observations made at the Observatorio del Roque de los Muchachos with the William Herschel Telescope (WHT) and the Gran Telescopio Canarias (GTC) operated on the island of La Palma by the IAC. This work is supported by the Spanish Ministry of Science under grants PID2020--120323GB--I00, EUR2021--122010 and AYA2017-83216-P. We acknowledge support from the ACIISI, Consejería de Economía, Conocimiento y Empleo del Gobierno de Canarias and the European Regional Development Fund (ERDF) under grant with reference ProID2021010132 and ProID2020010104. TMD and MAPT acknowledge support via Ram\'on y Cajal Fellowships RYC--2015--18148 and RYC--2015--17854, respectively.

\section*{Data Availability}

The GTC and WHT data are publicly available from \url{https://gtc.sdc.cab.inta-csic.es/gtc/jsp/searchform.jsp} and \url{http://casu.ast.cam.ac.uk/casuadc/ingarch/query}, respectively.

\bibliographystyle{mnras}
\bibliography{references} 


\bsp	
\label{lastpage}
\end{document}